\documentclass[twocolumn, prl, superscriptaddress, notitlepage]{revtex4-1}
\usepackage{graphicx}
\usepackage{physics}
\usepackage{color}
\usepackage{amsmath}
\usepackage{amsfonts}
\usepackage{xspace}
\usepackage[dvipsnames]{xcolor}
\usepackage[breaklinks,colorlinks,linkcolor=blue,citecolor=blue,urlcolor=blue]{hyperref}

\newcommand{\Poincare}{Poincar\'e\xspace}

\begin{document}
\title{$SU(1,1)$ echoes for breathers in quantum gases}

\author{Chenwei Lv}
\thanks{They contribute equally to this work.}
\affiliation{Department of Physics and Astronomy, Purdue University, West Lafayette, IN, 47907, USA}

\author{Ren Zhang}
\thanks{They contribute equally to this work.}
\affiliation{Department of Physics and Astronomy, Purdue University, West Lafayette, IN, 47907, USA}
\affiliation{School of Science, Xi'an Jiaotong University, Xi'an, Shaanxi 710049}

\author{Qi Zhou}
\email{zhou753@purdue.edu}
\affiliation{Department of Physics and Astronomy, Purdue University, West Lafayette, IN, 47907, USA}
\affiliation{Purdue Quantum Science and Engineering Institute, Purdue University, West Lafayette, IN 47907, USA}
\date{\today}
\begin{abstract}
Though the celebrated spin echoes have been widely used to reverse quantum dynamics, 
they are not applicable to systems whose constituents are beyond the control of the $su(2)$ algebra. 
Here, we design echoes to reverse quantum dynamics of breathers in three-dimensional unitary fermions 
and two-dimensional bosons and fermions with contact interactions, which are governed 
by an underlying $su(1,1)$ algebra.
Geometrically, $SU(1,1)$ echoes produce closed trajectories on a single or multiple \Poincare disks 
and thus could recover any initial states without changing the sign of the Hamiltonian. 
In particular, the initial shape of a breather determines the superposition of trajectories 
on multiple \Poincare disks and whether the revival time has period multiplication. 
Our work provides physicists with a recipe to tailor collective excitations 
of interacting many-body systems.
\end{abstract}
\maketitle 

It is notoriously difficult to reverse quantum many-body dynamics, 
which requires changing the signs of all terms in the Hamiltonian simultaneously 
and reversing the dynamics of all particles in a synchronized means.
Nevertheless, the well-established spin echoes \cite{Hahn1} 
have been widely used to overcome dephasing in spin systems, 
laying the foundation of many modern technologies, 
ranging from the nuclear magnetic resonance to the central spin problem in 
condensed matter systems \cite{Hahn2,Stejskal1,Liu1}.

The study of collective excitations has been a main theme in 
ultracold atoms and related topics \cite{bogoliubov1947,Stringari1,Matthews1,Gritsev1,Pethick1}.
Breathing modes (or breathers) of interacting fermions and bosons 
have provided physicists with valuable information about superfluidity and hydrodynamics 
in the past two decades \cite{Griffin1,Heiselberg1,Kinast1,Turlapov2007,Kinast2,Altmeyer1}. 
However, it is in general a grand challenge to recover the initial state once collective excitations are generated. 
The standard spin echoes do not apply to these breathers, whose relevant degrees of freedom 
do not obey the $su(2)$ algebra. 
A crucial question then arises.
Could we reverse many-body dynamics of breathers in interacting bosons and fermions? 

In this work, we implement the $SU(1,1)$ group
to design echoes to reverse collective excitations of quantum gases.  
If the initial state is an eigenstate of a harmonic trap, 
$SU(1,1)$ echoes can be geometrized using a single \Poincare disk and guarantee that the initial state returns at $2nT$, 
where $n$ is an integer and $T$ is the period of repeated drivings. 
When the initial state is not an eigenstate of a harmonic trap, multiple \Poincare disks are required to describe the dynamics. 
The interference between trajectories on these \Poincare disks determines whether the revival time is $2T$ or longer, 
the latter corresponding to period multiplication. 
When incommensurate frequencies exist in the dynamics, the revival time extends to infinity. 
These results shed light on remarkable phenomena observed in a recent experiment by Dalibard's group at ENS \cite{Dalibard1}. 

Following the seminal work by Pitaevskii and Rosch \cite{Pitaevskii1},  
breathers in quantum gases have been extensively studied \cite{anomaly1,anomaly2,anomaly3,anomaly4,anomaly5,Hofmann,Kagan1, Chevy1, Vogt1}. 
However, the fundamentally important role of initial shapes was not noted until the ENS experiment \cite{Dalibard1}, 
which found that the period of a triangular breather agrees  
with well-known results in harmonic traps 
when the quantum anomaly is negligible. 
In sharp contrast, an initial disk shape leads to an unprecedented period multiplication, 
quadrupling that of a triangle.  Such an observation is readily beyond understandings built upon previous works 
\cite{Pitaevskii1,anomaly1,anomaly2,anomaly3,anomaly4,anomaly5,Hofmann,Kagan1, Chevy1, Vogt1}.
More strikingly, other shapes 
do not have regular periodicities in experimentally accessible timescales, 
though the underlying Hamiltonian of breathers naturally defines a period. 
This remarkable ENS experiment remains unexplained as of now. Here, we show that these 
extraordinary behaviors of breathers originate from an intrinsic property of representing the $SU(1,1)$ group. 
In particular,
the underlying algebra and the geometric representation of $SU(1,1)$ echoes 
allow us to infer how initial shapes of breathers lead to distinct superpositions of \Poincare disks and consequently, the revival times.

\begin{figure}[tp] 
  \includegraphics [width=0.48\textwidth]{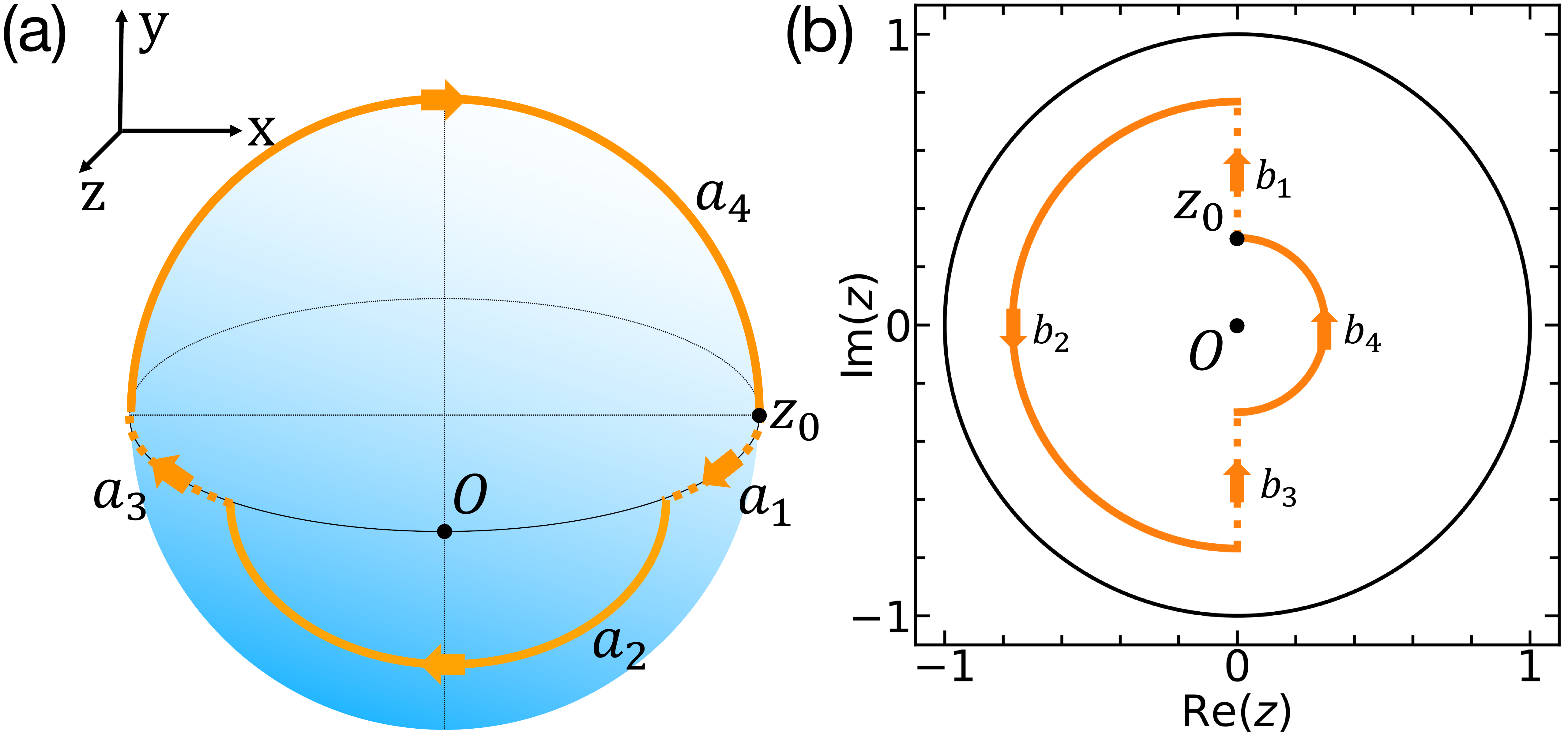}
  \caption
  {(a) A spin echo on a Bloch sphere. 
  $z_0$ denotes the initial state. 
  $a_{1}$ and $a_{3}$ represent rotations about the $y$ axis. 
  $a_{2}$ and $a_{4}$ represent $\pi$-pulses about the $z$ axis.  
  (b) A $SU(1,1)$ echo on a \Poincare disk.  
  $b_{1}$ and $b_{3}$ represent boosts, which are induced by the same Hamiltonian, $H$, along a radial direction. 
  $b_{2}$ and $b_{4}$ represent $\pi$-rotations about the origin.  }
  \label{fig:fig1}
\end{figure}

Generators of $SU(1,1)$ satisfy
\begin{equation}\label{alg}
    [K_1,K_2]=-iK_0,\ [K_2,K_0]=iK_1,\ [K_0,K_1]=iK_2.
\end{equation}
Geometrically, $SU(1,1)/U(1)$ corresponds to a \Poincare disk \cite{Novaes2004}, 
where each point on the disk is a $SU(1,1)$ coherent state, as shown in Fig. \ref{fig:fig1}. 
Such a coherent state characterized by a complex number $|z|<1$ is written as 
\begin{equation}
\ket{k,z}=(1-|z|^2)^k\sum_{n=0}^\infty \sqrt{\frac{\Gamma(2k+n)}{\Gamma(n+1)\Gamma(2k)}} z^n \ket{k,n},
\end{equation}
where $\Gamma(x)$ is the gamma function.
$k$ is determined by the Casimir operator $C=K_0^2-K_1^2-K_2^2$, $C\ket{k,n}=k(k-1)\ket{k,n}$. 
A single \Poincare disk is characterized by a unique $k$. 
$n$ is obtained from $K_{0}\ket{k,n}=(k+n)\ket{k,n}$.

$SU(1,1)$ has been widely applied in multiple disciplines 
\cite{Quesne1,Bander1,Romero1,Aravind1,Wodkiewicz1,Yurke1,Gerry1,Chiao1,Castin1, Son1,Deng371,Elliott1,Campo1}. 
However, it has not been implemented to study echoes until very recently. 
We have found that the dynamical instability of a BEC induced by quenching the scattering length,
which corresponds to a particular realization of the $SU(1,1)$ group, 
could be reversed by a family of $SU(1,1)$ echoes \cite{Lyu1}. 
The same representation of the $SU(1,1)$ group has also been considered for studying periodically driven BECs in \cite{Zhai1,zheyu}. 
In such a particular representation \cite{Lyu1,Zhai1,zheyu}, 
$k$ is either a positive integer or half integer. 
This is similar to spin systems, 
whose Casimir operator is equivalent to the angular momentum. In both cases, the integral or half-integral $k$ guarantees 
an echo has a single period. 
In sharp contrast, breathers considered here correspond to a distinct representation that has a continuous spectrum of $k$. 
Such fundamental difference provides breathers with much richer phenomena 
ranging from an arbitrary multiplication of the period to dynamics with non-commensurate frequencies.

$K_0$ is the Hamiltonian of trapped BECs \cite{Pitaevskii1},
\begin{equation}\label{gen}
\begin{split}
  K_0&=\frac1{2}\left[\sum\nolimits_i-\frac{1}{2}\nabla_i^2+\frac1{2}r_i^2+\sum\nolimits_{i< j}V({\bf r}_i-{\bf r}_j)\right],\\
  K_1&=\frac1{2}\left[\sum\nolimits_i-\frac{1}{2}\nabla_i^2-\frac1{2}r_i^2+\sum\nolimits_{i< j}V({\bf r}_i-{\bf r}_j)\right],\\
  K_2&=\frac1{4i}\sum\nolimits_i({\bf r}_i\cdot\nabla_i+\nabla_i\cdot{\bf r}_i).\\
\end{split}
\end{equation}
We have chosen $l_{ho}=\sqrt{\hbar/(m\omega_0)}$ as the unit length 
and $\hbar \omega_0$ as the unit energy. 
In three dimensions, both non-interacting systems and unitary fermions satisfy the commutators in Eq.(\ref{alg}).  
In the latter case, $V({\bf r}_i-{\bf r}_j)$ should be understood as 
$V({\bf r}_i-{\bf r}_j)=\tilde{V}({\bf r}_i-{\bf r}_j)\delta_{\sigma_i\neq\sigma_j}$, where
$\sigma_i=\uparrow, \downarrow$, and produces a divergent scattering length. 
In two dimensions, fermions and bosons with contact interactions, 
$V({\bf r}_i-{\bf r}_j)\sim g\delta({\bf r}_i-{\bf r}_j)$, also have the $SU(1,1)$ symmetry, 
as $\delta_{2D}(\lambda {\bf r})=\lambda^{-2}\delta_{2D}({\bf r})$. 
In single-component bosons, interactions exist between any pair of particles. 

\begin{figure*}[htp!] 
  \includegraphics [angle=0,width=.95\textwidth]
  {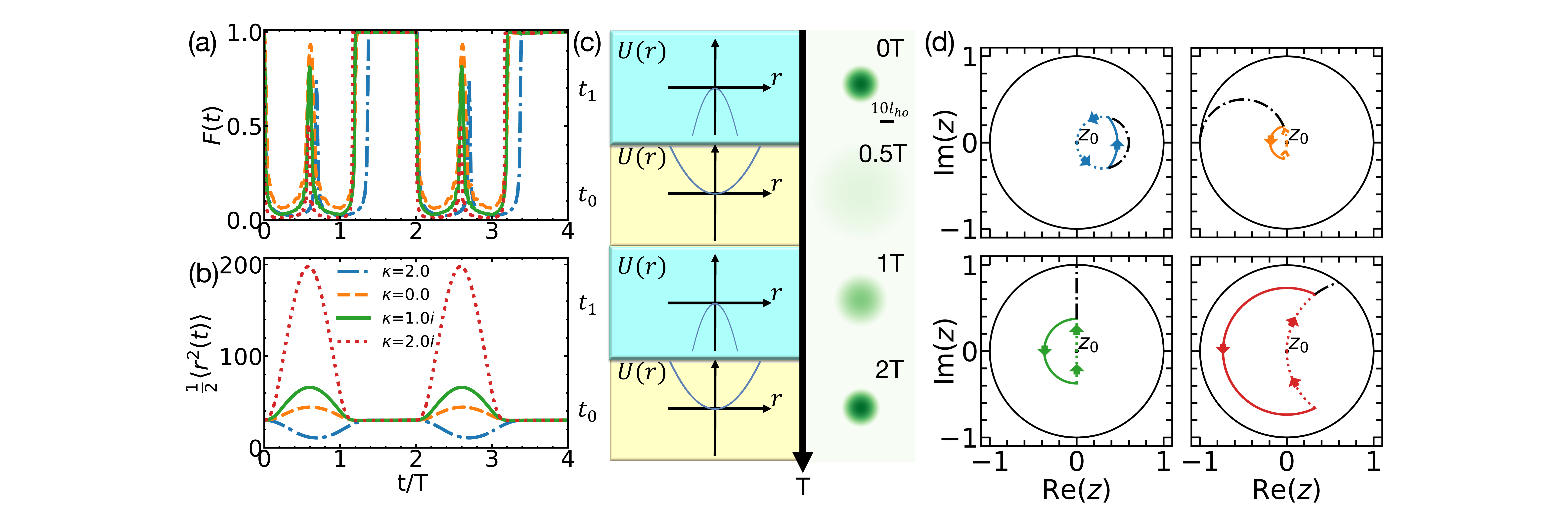}
  \caption
  { (a-b) $F(t)$ and $\langle r^2(t)\rangle/2$ of 2D BECs. 
  The initial state is the ground state of $K_0$. 
  $\kappa=2,0$ correspond to a modified and vanishing harmonic trap in the time interval, $nT<t<nT+t_1$, respectively.  
  $\kappa=i,2i$ correspond to inverted harmonic traps.
  $Ng=25600$, $\omega_0=20\times 2\pi {\rm Hz}$ and $t_1=\pi/8$. $t_0$ is determined by Eq.\ref{ect2}. 
  (c) Left panel: harmonic traps in different time intervals. 
  Right panel: snapshots of densities at different times for $\kappa=2i$. 
  (d) Trajectories on the \Poincare disk. 
  Dotted and solid lines are evolutions governed by $H_1$ and $H_0$, respectively. 
  Dot-dashed lines show the trajectories if only $H_1$ is applied. 
  }\label{fig:fig2}
\end{figure*}

$SU(1,1)$ echoes arise from the identity, 
\begin{equation}
e^{-i(\varphi_1K_1+\varphi_2K_2)}e^{-i\pi K_0}e^{-i(\varphi_1K_1+\varphi_2K_2)}e^{i\pi K_0}=\mathcal{I},\label{echo}
\end{equation}
where $\mathcal{I}$ is the identity operator, $\varphi_1$ and $\varphi_2$ are two arbitrary real numbers.  
On the \Poincare disk, $e^{-i\pi K_0}$ is a rotation of $\pi$ about the origin,
and $e^{-i(\varphi_1K_1+\varphi_2K_2)}$ is a boost changing $|z|$.
A simple echo is illustrated in Fig.\ref{fig:fig1}(b).
Starting from a given initial state, $\mathcal{U}_1=e^{-i\varphi_1K_1}$ moves it
along a diameter and is followed by a rotation $\mathcal{U}_0=e^{-i\pi K_0}$. 
Using Eq.(\ref{echo}), we conclude $(\mathcal{U}_0\mathcal{U}_1)^2=e^{-i2\pi K_0}$, 
i.e., a rotation of $2\pi$ about the origin, and thus the initial state is recovered. 
This echo applies to any initial states on the \Poincare disk and any $\varphi_1K_1+\varphi_2K_2$. 

Whereas our results apply to any eigenstates of a harmonic trap,
we first choose the ground state of the Hamiltonian, $H_0=2K_0$,
as the initial state as an example to demonstrate our scheme.
To implement a $SU(1,1)$ echo, the trapping frequency is suddenly changed to $\omega_1=\kappa \omega_0$ at $t=0$, 
where $\kappa$ is an arbitrary real or imaginary number. 
In the latter case, it corresponds to an inverted harmonic trap. 
When $t=t_1$, the original harmonic trap is restored and the system evolves for another time period $t_0$. 
Then the above two steps are repeated. 
Such dynamics are governed by the Hamiltonians, 
\begin{align}
  H_1&=(1+\kappa^2)K_0+(1-\kappa^2)K_1,\quad nT<t<nT + t_1,\nonumber\\
  H_0&=2K_0, \qquad\,\,nT + t_1<t<(n+1)T, \label{H}
\end{align}
where $n$ is a {non-negative} integer, and $T=t_0+t_1$ defines a period. 
The propagator, $(\mathcal{U}_0\mathcal{U}_1)^2=(e^{-iH_0t_0}e^{-iH_1t_1})^2$, 
from $t=nT$ to $t=(n+2)T$ can be rewritten as
\begin{equation}
  e^{-i(\zeta_1+2t_0)K_0}e^{-i\eta_1K_1}e^{-i(2\zeta_1+2t_0)K_0}e^{-i\eta_1K_1}e^{-i\zeta_1K_0},
\end{equation}
where $\zeta_1=\arctan(\frac{1+\kappa^2}{2\kappa}\tan\kappa t_1)$ 
and $\eta_1=2{\rm arcsinh}\left(\frac{1-\kappa^2}{2\kappa}\sin(\kappa t_1)\right)$.
We have used the Baker-Campbell-Hausdorff decomposition,
\begin{equation}
e^{-i(\xi_0K_0+\xi_1K_1+\xi_2K_2)}= e^{-i\zeta K_0}e^{-i\eta(K_1\cos\phi+K_2\sin\phi)}e^{-i\zeta K_0},
\end{equation}
where
$\tan\zeta=(\frac{\xi_0}{\xi}\tan\frac{\xi}{2})$, $\cos\phi=(\frac{\xi_1}{\sqrt{\xi_1^2+\xi_2^2}})$, 
$\sinh(\frac{\eta}{2})=\frac{\sqrt{\xi_1^2+\xi_2^2}}{\xi}\sin(\frac{\xi}{2})$, and $\xi^2=\xi_0^2-\xi_1^2-\xi_2^2$. 
To deliver an echo, it is required that $\pi=2(t_0+\zeta_1)$, or equivalently, 
\begin{equation}\label{ect2}
t_0=\frac{\pi}{2}-\zeta_1=\frac{\pi}{2}-\arctan(\frac{1+\kappa^2}{2\kappa}\tan\kappa t_1). 
\end{equation}
Under this condition, $(\mathcal{U}_0\mathcal{U}_1)^2=e^{-i2\pi K_0}$.
As any eigenstate of a harmonic trap is also eigenstate of $C$ with an eigenvalue $k(k-1)$, 
only an extra overall phase shows up and the system returns to the initial state after two driving periods.
Once $H_0$ and $H_1$ are fixed, 
tuning $t_0$ to deliver a $SU(1,1)$ echo is an analogue of adjusting the duration of the pulse to create a $\pi$ rotation on a Bloch sphere in spin echoes.
There also exist echoes allowing the initial state to return in a longer time, 
say $t=3nT$ (Supplementary Materials).

The expectation value of the potential energy, 
$E_{\text{pot}}=\langle\frac{1}{2} \sum_ir_i^2\rangle$ in the time interval $nT+t_1<t<(n+1)T$
can be written as $E_{\text{pot}}=\langle K_0-K_1\rangle$. 
Using properties of $SU(1,1)$ coherent states, 
$\langle k,z|K_0|k,z\rangle=k\frac{1+|z|^2}{1-|z|^2}$, $
\langle k,z|K_1|k,z\rangle=2k\frac{{\rm Re}(z)}{1-|z|^2}$, 
we obtain, 
\begin{equation}
E_{\text{pot}}=k[{1+|z|^2-2{\rm Re}(z)}]/({1-|z|^2}). \label{Ep}
\end{equation}
$E_{\text{pot}}$ in the time interval $nT<t<nT+t_1$ simply multiplies the above equation by $\kappa^2$. 
Apparently, $E_{\text{pot}}$ is periodic with period $2T$. 
For the system prepared in the ground state of $H_0$ with ground state energy $E_g$, we have $k=E_g/2$.
Results above are valid for any eigenstates of the initial Hamiltonian, 
hold for any finite temperatures at thermal equilibrium, 
and {$k$ in} Eq. (\ref{Ep}) should be understood as
$\langle K_0\rangle_{\rm thermal}={\rm Tr}(K_0 e^{-\beta H_0})/{\rm Tr}( e^{-\beta H_0})$, 
where $\beta$ is the inverse temperature.

It is useful to consider 2D bosons as an example. 
Whereas it is difficult to compute the exact many-body state in the quantum dynamics controlled by the $su(1,1)$ algebra, 
in the weakly interacting regime, such dynamics is well captured by a Gross-Pitaevskii (GP) equation,
\begin{equation}
  i\frac{\partial \Psi({\bf r},t)}{\partial t}=\left(-\frac{\nabla^2}{2}+\frac{\kappa(t)^2 r^2}{2}+gN|\Psi({\bf r},t)|^2\right)\Psi({\bf r},t),
 \end{equation}
where $N$ is the number of bosons, $g=4\pi a_0$ with $a_0$ being the dimensionless scattering length. 
We use an imaginary time evolution to obtain the ground state of the initial Hamiltonian, $H_0$. 
We then let the condensate evolve based on the GP equation, 
in which the Hamiltonian is determined by Eq.(\ref{H}). 
We trace both the overlap between $\Psi({\bf r},t)$ and $\Psi({\bf r},0)$, 
$F(t)=|\int d{\bf r} \Psi^*({\bf r},0)\Psi({\bf r},t)|$, 
and the absolute value of the potential energy, $|E_{\text{pot}}|$. 
Fig. \ref{fig:fig2} shows a few typical choices. 
(I), $\kappa$ is real, corresponding to a harmonic trap 
whose frequency could be different from the initial one.
(II), $\kappa=0$, corresponding to turning off the harmonic trap. 
(III), $\kappa$ is purely imaginary, meaning an inverted harmonic trap. 
For a generic $H=\sum_{i=0,1,2} \xi_i K_i$,
$\vec{\xi}=\{\xi_0,\xi_1,\xi_2\}$ defines an external field with a strength, $\xi^2=\xi_0^2-\xi_1^2-\xi_2^2$.
For instance, in Eq.(\ref{H}), we have $\xi=2\kappa$.
In (I), $\xi^2>0$, and the system follows a closed loop on the \Poincare disk. 
In (II), $\xi$ vanishes. Without a confining potential in the real space, 
the trajectory on the \Poincare disk eventually becomes tangent with the boundary circle. 
In (III), $\xi$ becomes purely imaginary. 
While a deconfining potential pushes BECs to expand in the real space, 
on the \Poincare disk, the trajectory becomes an open path.
Though quantum dynamics governed by $H_1$ alone in (I-III) are distinct, 
$SU(1,1)$ echoes always lead to revivals. 
Fig. \ref{fig:fig2} clearly shows that both $F(t)$ and $|E_{\text{pot}}|$ 
are periodic functions of $t$ with a period of $2T$. 

\begin{figure}[tp] 
  \includegraphics [angle=0,width=.5\textwidth]
  {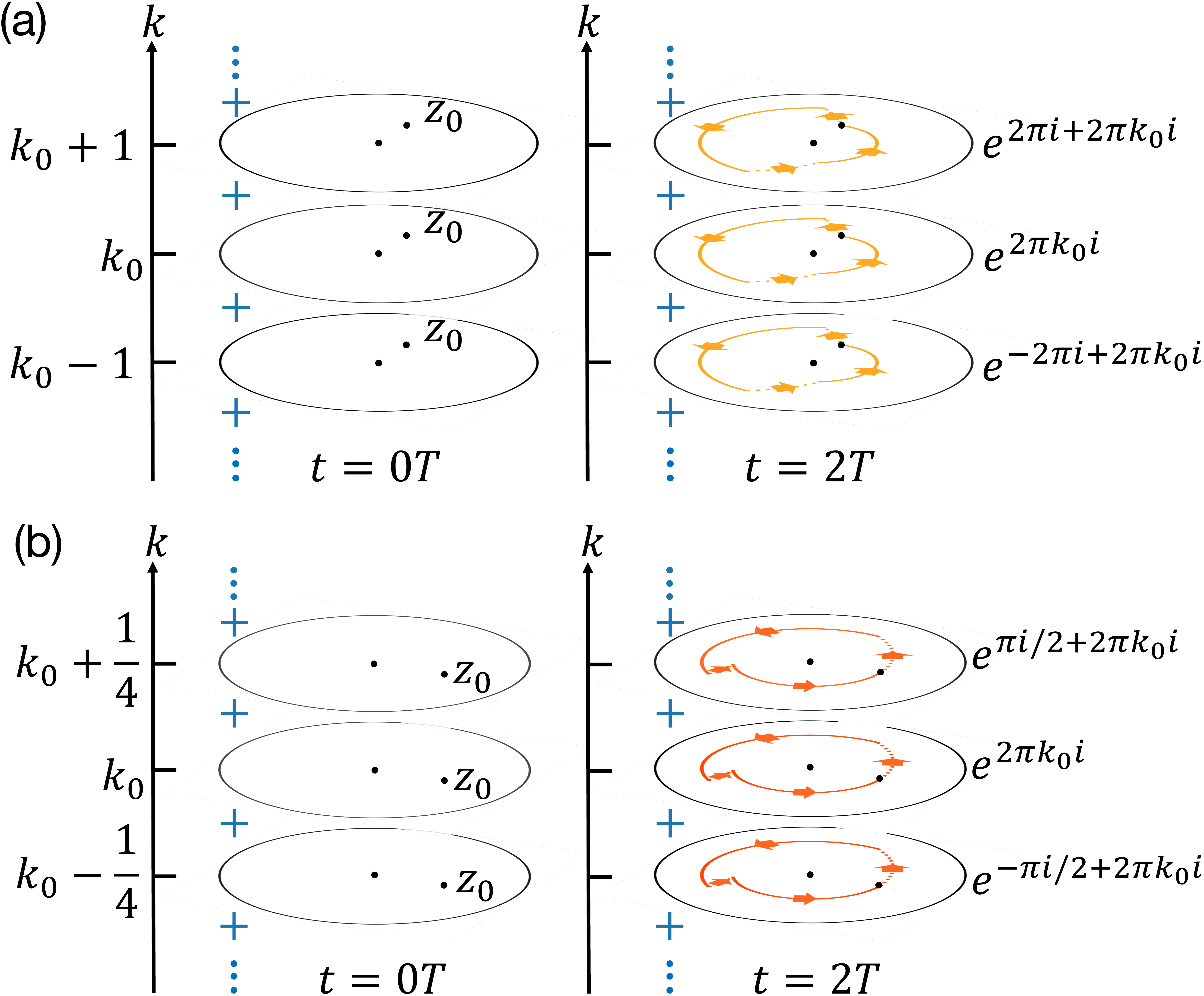}
  \caption
  {
  (a) At $t=2T$, trajectories on different disks accumulate the same phase. 
  The system returns to the initial state. 
  (b) Trajectories on different disks acquire relative phases. 
  It takes the system $8T$ to return to the initial state.
  }\label{fig:fig3}
\end{figure}

The initial state could also be a superposition of multiple eigenstates of $C$
such that multiple \Poincare disks are required.
We consider an arbitrary propagator $\mathcal{U}$ in the $SU(1,1)$ group acting on 
$\ket{\Psi}=\sum_{n,k} c_{nk}\ket{k,n}=\sum_k \ket{\psi_k}$, where $\ket{\psi_k}=\sum_n c_{nk}\ket{k,n}$ and 
$\langle\psi_{k'}| \psi_k\rangle\sim \delta_{k,k'}$. 
Here we have suppressed other quantum numbers for the same {$k,n$}. 
As $\mathcal{U}{\ket{\Psi}}=\sum_{k} \mathcal{U}\ket{\psi_k}$, 
and $\mathcal{U}\ket{\psi_k }$ corresponds to an evolution on a single \Poincare disk, 
the dynamics thus correspond to superpositions of trajectories on multiple \Poincare disks. 
If an echo, $(\mathcal{U}_0\mathcal{U}_1)^2=e^{-2i\pi K_0}$, acts on the initial state for $m$ times, 
where $m$ is an integer, we obtain
\begin{equation}\label{echoevo}
  e^{-2i\pi m K_0}\ket{\Psi}=\sum\nolimits_{n,k} c_{nk}e^{-2i\pi m k}\ket{k,n}.
\end{equation}
Replacing the sum in Eq.(\ref{echoevo}) by an integral over $k$, 
this initial state shall, in general, 
include incommensurate $k$'s and thus lacks a finite periodicity. 
Here we consider systems with well defined periodicities 
and apply summation of discrete $k$'s.
Since $e^{-2i\pi m k}$ is independent of $n$, 
the return probability $P(m)=|\langle\Psi(0)|\Psi(2mT)\rangle|^2$ becomes 
$P(m)=\left|\sum_{k} \tilde{P}_k e^{-2i\pi m k}\right|^2$, 
where $\tilde{P}_k=\sum\nolimits_n |c_{nk}|^2$, $\sum_k \tilde{P}_k=1$. 
It is apparent that $P(m)=1$ only if 
 $ e^{-2i\pi m k}=e^{i\phi_0}$ 
for all $k$'s with a nonzero $c_{nk}$,
where $\phi_0\in[0,2\pi)$ is independent of $k$. 
This is certainly satisfied if the initial state includes a single state, $\ket{k_0,n_0}$. 
It is also clear that $e^{-2i\pi m k}=e^{i\phi_0}$
will never be satisfied if $\ket{\Psi}$ includes incommensurate $k$'s. 
Whereas such a scenario is impossible in previous works
\cite{Lyu1,Zhai1,zheyu}, 
in breathers with a continuous spectrum of $k$, dynamics controlled by incommensurate $k$'s may arise.

If $k$'s in Eq.(\ref{echoevo}) are commensurate, i.e.,  
all $k$'s are represented by $k=k_0+p/Q$, 
where $k_0$ is a given reference with a nonzero $c_{nk_0}$, $p\in \mathbb{Z}$, $Q\in\mathbb{N}_+$, 
and $p$ and $Q$ are co-prime numbers, we have $P(Q)=1$, 
and the system evolves back to its initial state after $2Q$ periods. 
Therefore, different superpositions of $|k,n\rangle$ 
in the initial state may lead to distinct revival times after $SU(1,1)$ echoes are applied. 
If $Q>1$, period multiplication emerges in the dynamics. 
Fig.\ref{fig:fig3} shows examples corresponding to $Q=1$ and $Q=4$. 
Consequently, revival times are $2T$ and $8T$, i.e., period quadruples in the latter case. 

\begin{figure}[tp] 
  \includegraphics [angle=0,width=.5\textwidth]
  {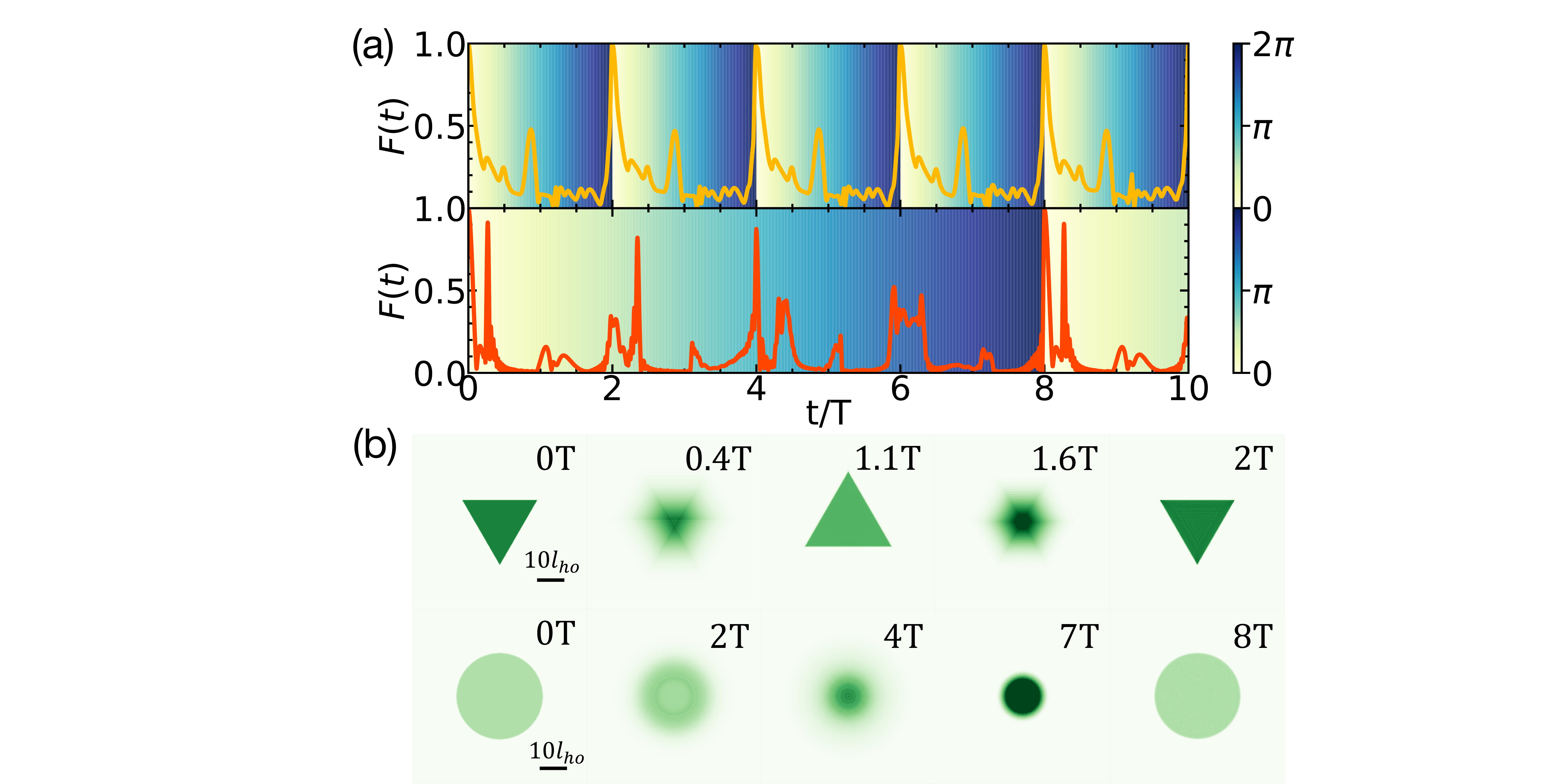}
  \caption
  {
  (a) $F(t)$ of breathers with an initial triangular (top) and disk (bottom) shapes, respectively. 
  $\omega_0=40\times 2\pi {\rm Hz}$, $t_1=\pi/8$, and $\kappa=0.5i$. 
  $Ng=25600$ ($12800$) is used for the triangle (disk).
  The background color represents the time-dependent relative phase between different \Poincare disk.
  (b) Density distributions of BECs at different times.   } \label{fig:fig4}
\end{figure}

Applying the above analysis to breathers of different initial shapes, 
we observe that the triangle and the disk correspond to 
$Q=1$ and $Q=4$, respectively. 
The initial state is chosen as the ground state of a flat-box potential with an infinite potential wall. 
Such an initial state is no longer an eigenstate of $K_0$, 
and it is useful to fully implement the $su(1,1)$ algebra to consider the dynamics.
After turning off the flat-box potential, the system evolves based on Eq.(\ref{H}). 
$F(t)$ of a triangle satisfies $F(t)=F(t+2T)$. 
As the initial state is not an eigenstate of $K_0$, 
it must be a superposition of multiple $|k,n\rangle$ with 
differences between $k$'s being integers, i.e., $Q=1$ as shown in
Fig.\ref{fig:fig3}(a). 
The exact number of \Poincare disks can be, in principle, determined 
by considering a particular Hamiltonian, 
$\tilde{H}\equiv C$, $e^{-i \tilde{H}t}\ket{\Psi}=\sum_{nk}c_{nk}e^{-ik(k-1)t}|n,k\rangle=\sum_{k}e^{-ik(k-1)t}|\psi_k\rangle$. 
A Fourier transform of $F(t)=\sum_k e^{-ik(k-1)t}\langle \psi_k |\psi_k\rangle$ to the frequency space unfolds how many $k$'s 
are involved and their corresponding weights. 
Nevertheless, such calculations are not essential here, since our echoes apply to any 
superpositions in Eq.(\ref{echoevo}), 
regardless of the exact number of \Poincare disks involved. 

The results of a disks shape are distinct. 
Fig.\ref{fig:fig4} shows that the revival time of the disk is $8T$. 
We conclude that the superposition in the initial state must be similar to Fig.\ref{fig:fig3}(b).  
The quench dynamics in the ENS experiment has a propagator, 
$e^{-iK_0t}$, corresponding to $SU(1,1)$ echoes where $t_{1}=0$.
Such quench dynamics has a periodicity of $2T$ and $8T$ for the triangle and the disk, respectively \cite{Dalibard1}. 
This also confirms that the triangle and the disk corresponds to a superposition of 
multiple \Poincare disks with $Q=1$ and $Q=4$, respectively.  We have not found other shapes, such as a square, 
which return to the initial states within timescales of our numerical simulations, 
similar to results of the quench dynamics \cite{Dalibard1}. 
We conclude that these shapes are described by either incommensurate $k$'s or commensurate $k$'s corresponding to a very large $Q$, 
which lead to revival times not observable in relevant timescales of numerics and experiments.

In experiments, it is the exact many-body state that evolves under the control of $SU(1,1)$ echoes. 
Results of the GP equation are expected to provide us with a good approximation in the weakly interacting limit. 
Nevertheless, the precise form of the many-body state corresponding to a given initial shape of the breather 
remains an interesting open question worthy of future studies. 
In contrast, $c_{nk}$ can be straightforwardly obtained in few-body systems. 
For instance, in a two-body problem, 
eigenvalues of the Casimir operator are directly related to the angular momenta 
such that the initial shape of the breather allows one to directly predict the
revival time (Supplementary Materials).  
Similar to spin echoes, $SU(1,1)$ echoes could be implemented to detect symmetry breaking perturbations, 
such as an extra external potential in experiments (Supplementary Materials).

Our results are obtained by an algebraic method independent on the representation 
and apply to any systems with the $SU(1,1)$ symmetry. 
We hope that our work will stimulate more interests from different disciplines to use 
geometric approaches to control quantum dynamics in few-body and many-body systems.

\begin{acknowledgments}
  QZ is grateful to Jean Dalibard for useful discussions at Sant Feliu that stimulated this work 
  and for his many inspiring questions during later communications. 
  This work is supported by DOE DE-SC0019202, W. M. Keck Foundation, and a seed grant from PQSEI. 
  RZ is supported by the National Key R$\&$D Program of China (Grant No. 2018YFA0307601), NSFC (Grant No.11804268).
\end{acknowledgments}


%
  
  \onecolumngrid

  \newpage

  \vspace{0.4in}

  \centerline{\bf\large Supplementary Materials for ``$SU(1,1)$ echoes for breathers in quantum gases"}

  \vspace{0.2in}
\subsection{Echoes with periodicity $3T$}
In the main text, we have discussed the ${\it su}(1,1)$ echoes with a periodicity of $2T$, 
which arise from the identity in Eq. (4) of the main text. 
Here, we consider echoes with a periodicity of $3T$ and prove the following identity. 
\begin{equation}
   (\mathcal{U}_0\mathcal{U}_1)^3=e^{-i2\pi K_0}.
\end{equation}

\begin{figure}[ht] 
   \includegraphics [angle=0,width=.25\textwidth]
   {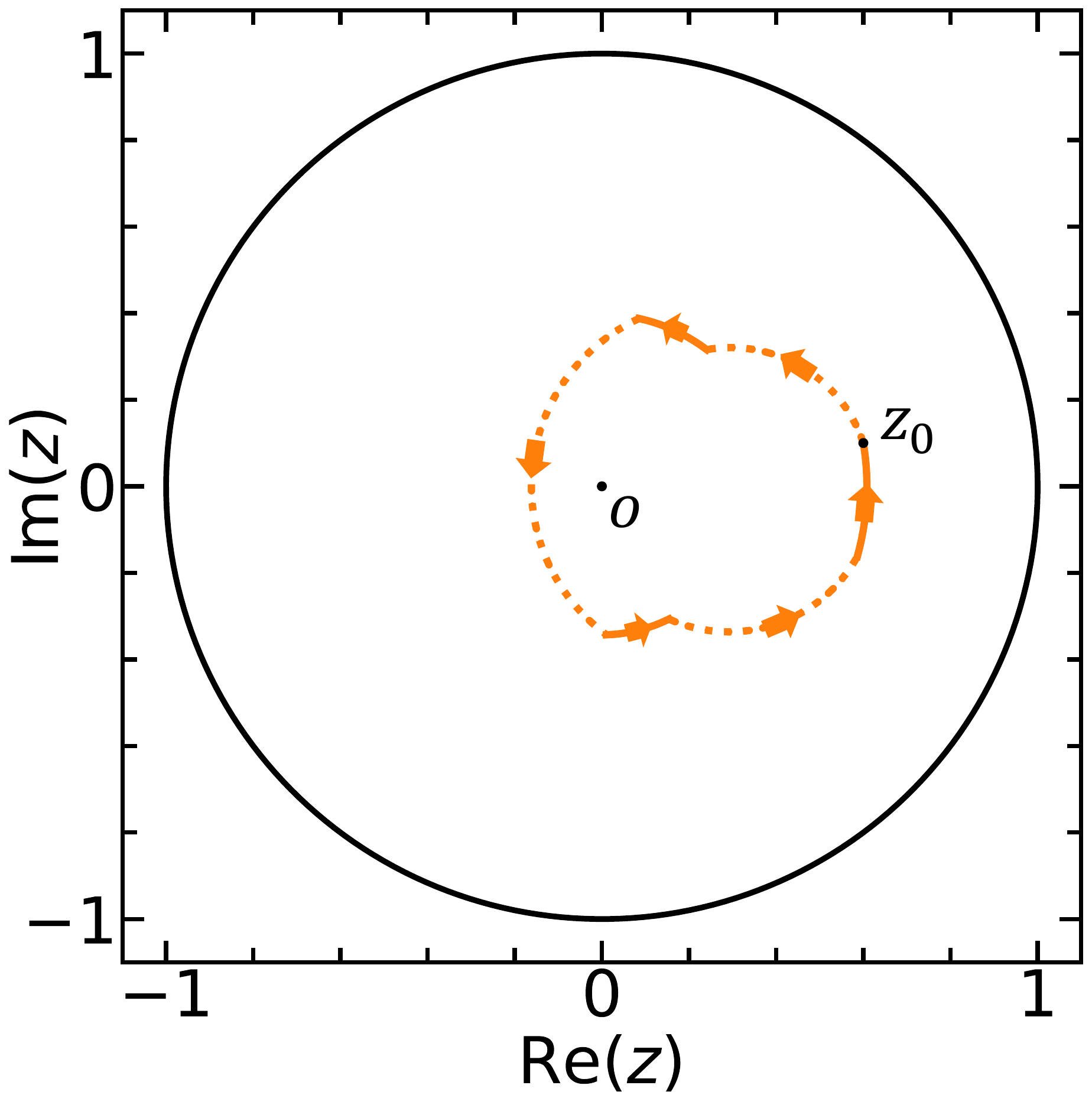}
   \caption
   {
      A trajectory on the Poincar\'e Disk, 
      where dotted and solid lines are evolutions governed by $H_1$ and $H_0$, respectively.
      We choose $\kappa=2$, $t_1=\pi/8$. $t_0$ is determined by Eq.\ref{ecthree}.
   }\label{figs1}
\end{figure}

\begin{figure}[ht] 
   \includegraphics [angle=0,width=.75\textwidth]
   {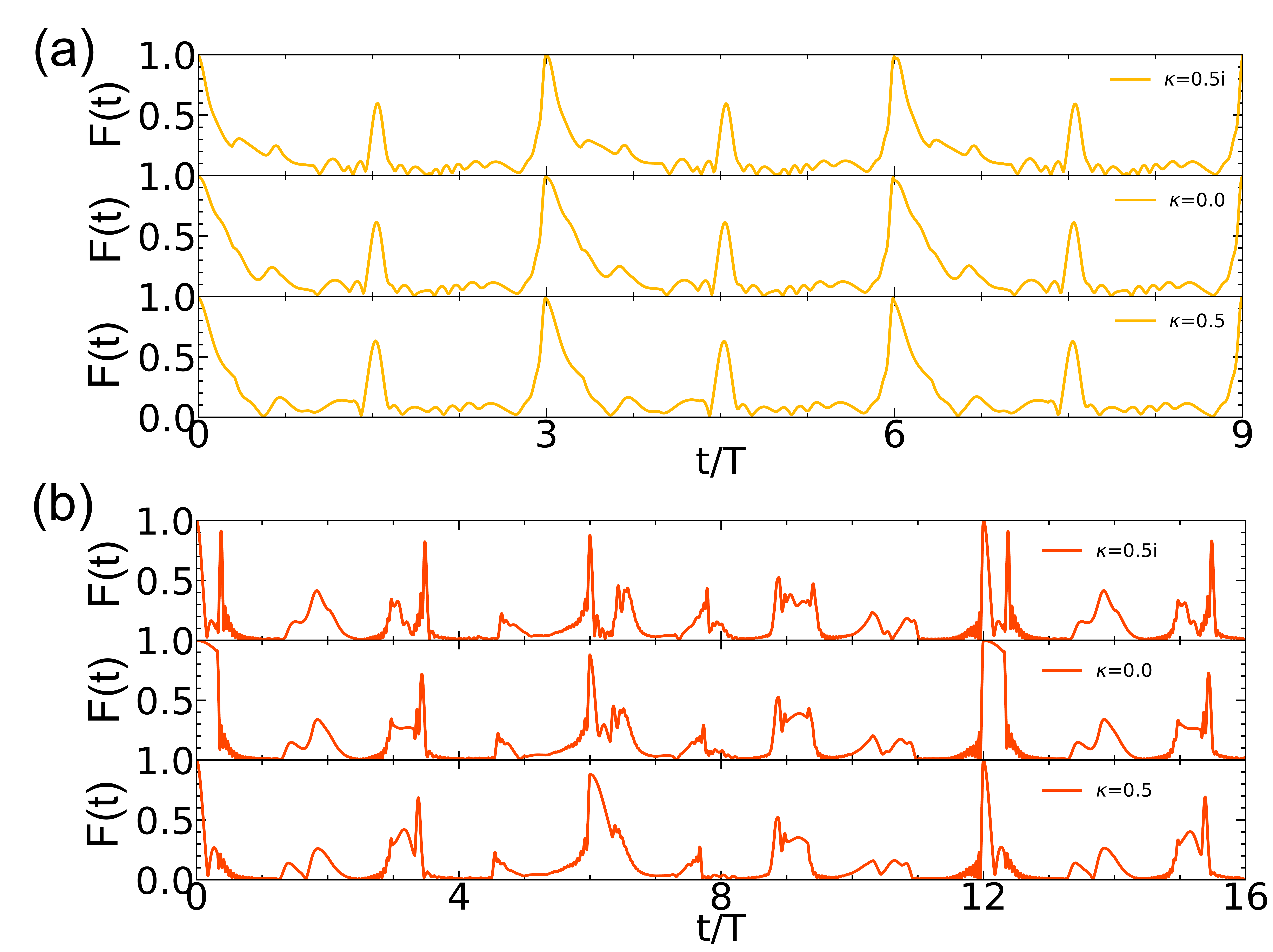}
   \caption
   { 
      (a)$F(t)$ for a triangle. $Ng=25600$, $\omega_0=40\times 2\pi {\rm Hz}$ and $t_1=\pi/8$.
      (b)$F(t)$ for a disk. $Ng=12800$, $\omega_0=40\times 2\pi {\rm Hz}$ and $t_1=\pi/8$. 
      For both cases $t_0$ is determined by Eq.\ref{ecthree}.
   }\label{figs2}
\end{figure}

We consider the same Floquet sequence as the main text, where the Hamiltonians controlling breathers in a harmonic trap 
are given by
\begin{align}
   H_1&=(1+\kappa^2)K_0+(1-\kappa^2)K_1,\quad nT<t<nT + t_1,\nonumber\\
   H_0&=2K_0, \qquad\,\,nT + t_1<t<(n+1)T, \label{H}
 \end{align}
 where $n$ is an integer, and $T=t_0+t_1$ defines a period. Using the BCH decomposition, we rewrite $(\mathcal{U}_0\mathcal{U}_1)^3$ as
\begin{equation}
   (\mathcal{U}_0\mathcal{U}_1)^3=e^{i\zeta_1K_0}\left(e^{-i\gamma K_0}e^{-i\eta_1K_1}\right)^3e^{-i\zeta_1K_0},
\end{equation}
where $\zeta_1=\arctan(\frac{1+\kappa^2}{2\kappa}\tan\kappa t_1)$, $\eta_1=2{\rm arcsinh}
\left(\frac{1-\kappa^2}{2\kappa}\sin(\kappa t_1)\right)$, and $\gamma=2\zeta_1+2t_0$.
To obtain
\begin{equation}\label{techo1}
   \left(e^{-i\gamma K_0}e^{-i\eta_1 K_1}\right)^3=e^{-i\alpha K_0},
\end{equation}
we adopt the $2\times 2$ representation of $su(1,1)$ algebra. Specifically, using Pauli matrices, we have
\begin{equation}
   K_0\to\sigma_z/2,\quad K_1\to i \sigma_x/2,\quad K_2\to i \sigma_y/2,
\end{equation}
and Eq.(\ref{techo1}) becomes
\begin{equation}\label{echoc}
   \begin{split}
      &\left(\cos\frac{\gamma}{2} \cosh\frac{\eta_1}{2} (f(\gamma,\eta_1)-2)\right)\\
      +&\left[\cos\frac{\gamma}{2}\sinh\frac{\eta_1}{2}\sigma_x+\sin(\frac{\gamma}{2})
      \sinh(\frac{\eta_1}{2})\sigma_y-i\cosh(\frac{\eta_1}{2})\sin(\frac{\gamma}{2})\sigma_z\right]
      f(\gamma,\eta_1)=\cos\frac{\alpha}{2}-i\sin\frac{\alpha}{2}\sigma_z,
   \end{split}
\end{equation}
where 
\begin{equation}
   f(\gamma,\eta_1)=\cos\gamma + (1 + \cos\gamma) \cosh\eta_1.
\end{equation}
Since $\cos(\gamma/2)$ and $\sin(\gamma/2)$ cannot vanish simultaneously, we require
$f(\gamma,\eta_1)=0$, which yields
\begin{equation}
   \gamma_0=\arccos(-\frac{\cosh\eta_1}{1+\cosh\eta_1}). \label{gamma0}
\end{equation}
Such $\gamma_0$ also satisfies $\cos\frac{\gamma_0}{2} \cosh\frac{\eta_1}{2}=\pm 1/2$. Thus, 
we obtain 
\begin{equation}
   \left[\left(\cos\frac{\gamma_0}{2}-i\sin\frac{\gamma_0}{2}\sigma_z\right)
   \left(\cosh\frac{\eta_1}{2}-\sinh\frac{\eta_1}{2}\sigma_x\right)\right]^3=-1=\cos\pi-i\sin\pi\sigma_z,
\end{equation}
and $\alpha=2\pi$. 
Therefore, we conclude that 
\begin{equation}
(\mathcal{U}_0\mathcal{U}_1)^3=e^{i\theta K_0}e^{-i2\pi K_0}e^{-i\theta K_0}=e^{-i2\pi K_0},
\end{equation}
if we choose
\begin{equation}\label{ecthree}
   t_0=\frac1{2}\arccos(-\frac{\cosh\eta_1}{1+\cosh\eta_1})-\zeta_1.
\end{equation}
We have verified these results by numerically solving the GP equation. 
As shown in Fig. \ref{figs1} and Fig. \ref{figs2},  the triangle and the disk need $3T$ and $12T$ to return to the initial state, respectively. 
Whereas echoes with even longer periods also exist, echoes of periods of $2T$ discussed in the main text are the simplest ones to implement in practice.

\subsection{Two unitary fermions in a harmonic trap}
We consider one spin-up and one spin-down fermion
in a three-dimensional harmonic trap, whose relative motion and the center of mass are decoupled. 
The ${\it su}$(1,1) algebra applies to both degrees of freedom.
The Hamiltonian is written as $H=H_{CM}+H_{rel}$,
\begin{equation}
   \begin{split}
      H_{CM}=&-\frac{\hbar^2}{2M}\nabla^2_{\bf R}+\frac1{2}M\omega_0^2R^2,\\
      H_{rel}=&-\frac{\hbar^2}{2\mu}\nabla^2_{\bf r}+\frac1{2}\mu\omega_0^2 r^2.
   \end{split}
\end{equation}
where $M=2m$, $\mu=m/2$, ${\bf R}=({\bf r}_1+{\bf r}_2)/2$, 
${\bf r}={\bf r}_1-{\bf r}_2$, $m$ is the mass of fermions. 
The interaction in the unitary limit is replaced 
by the boundary condition of the relative motion, namely
\begin{equation}
   \Psi_{rel}({\bf r})\underset{r\to 0}{\propto} \frac1{r},
\end{equation}
$\Psi_{rel}({\bf r})$ is the relative wave function and 
$\Psi({\bf R},{\bf r})=\Psi_{CM}({\bf R})\Psi_{rel}({\bf r})$.
Since the center of mass has the same simple dynamics of a single particle in a harmonic trap, 
we focus on the relative motion, 
which has the $SU$(1,1) generators \cite{Castins1}, 
\begin{equation}\label{gene2}
     K^{r}_0=\frac1{2}\left(-\frac{1}{2}\nabla^2+\frac1{2}r^2\right),
     \quad K^{r}_1=\frac1{2} \left(-\frac{1}{2}\nabla^2-\frac1{2}r^2\right),
     \quad K^{r}_2=\frac1{4i}\left({\bf r}\cdot\nabla+\nabla\cdot{\bf r}\right). 
\end{equation}
We have chosen the harmonic length $l_{ho}=\sqrt{\hbar/(\mu\omega_0)}$ 
as the unit length and $\hbar \omega_0$ as the unit energy. 
We first evaluate the Casimir operator of the ${\it su}$(1,1) algebra, 
\begin{equation}
   C=(K^{r}_0)^2-(K^{r}_1)^2-(K^{r}_2)^2=\frac{L^2}{4\hbar^2}-\frac3{16},
\end{equation}
where $L$ is the angular momentum operator for the relative motion. 
$C$ has the eigenvalue $l(l+1)/4-3/16$ for the angular momentum eigenstate $\ket{l,m}$.
Therefore we denote the eigenstate of $C$ by $\ket{k,n;l,m}$. 
Here $n$ is the principle quantum number, 
$k$ is the Casimir quantum number, 
$l$ and $m$ are the angular momentum and magnetic quantum number, respectively.
Since $C\ket{k,n;l,m}=k(k-1)\ket{k,n;l,m}=(l(l+1)/4-3/16)\ket{k,n;l,m}$, we obtain
\begin{equation}
   k=\frac{l}{2}+\frac3{4}, \quad l\geq 0;\qquad k=\frac1{4}\quad {\rm or}\quad  \frac3{4}, \quad l=0,
\end{equation}
where we only consider the positive discrete series for the representation of $SU$(1,1). 
Using $K_-\ket{k,0;0,0}=0$, we find 
\begin{equation}
   \bra{\bf r}\ket{\frac1{4},0;0,0}\underset{r\to 0}{\propto} \frac1{r},\qquad
   \bra{\bf r}\ket{\frac3{4},0;0,0}\underset{r\to 0}{\propto} 1.
\end{equation}
We conclude that the ground state of the relative motion of 
two unitary fermions in the $s$-wave channel corresponds to  $\ket{1/4,n;0,0}$. 
As for the center of mass, 
the same argument leads to $\ket{3/4,n;0,0}$. 
The spectrum of the relative motion becomes
\begin{equation}
   E_r=l+\frac3{2}+2n,\quad l\geq 1,\qquad E_r=\frac1{2}+2n,\quad l=0.
\end{equation}

The dynamics on multiple Poincar\'e disks can be generated by 
choosing an initial state as a superposition of different angular momentums.
For example, we consider the initial state as a mixture of $s$ and $d$-wave 
for their relative motion. 
The $s$-wave subspace has $k_s=1/4$ and the $d$-wave subspace has $k_d=7/4$ 
for any magnetic quantum number $m'$. 
The initial state and the state at the end of $2m$ Floquet periods are
\begin{equation}\label{eq:2uht}
  \begin{split}
    \ket{\Psi(0)}=&\ket{\Psi_{CM}}\otimes\sum\nolimits_n\left(s_n\ket{1/4,n;0,0}
    +\sum\nolimits_{m'}d_{nm'}\ket{7/4,n;2,m'}\right),\\
    \ket{\Psi(2mT)}=&\ket{\Psi_{CM}}\otimes\sum\nolimits_n\left(e^{-i(1/4+n)2\pi m}s_n\ket{1/4,n;0,0}+
    e^{-i(7/4+n)2\pi m}\sum\nolimits_{m'}d_{nm'}\ket{7/4,n;2,m'}\right), 
  \end{split}
\end{equation}
respectively. 
Since $k_s=k_d-3/2$, $Q=2$, using Eq. (\ref{eq:2uht}), 
we conclude that it takes $4$ Floquet periods for the systems to recover its initial state.

If the $s$-wave scattering length vanishes, the $s$-wave subspace has $k'_s=3/4$, 
while $d$-wave subspace still has $k'_d=7/4$. 
Therefore, we obtain $k'_s=k'_d-1$, $Q=1$. 
It takes the same initial state $2$ Floquet periods to return to the initial state.

\subsection{Using echoes to detect a $r^4$ potential}
As we have shown in Eq.(5-8) in the main text, we have considered systems with the
$SU$(1,1) symmetry and designed echoes using the $su$(1,1) algebra. Once a perturbation breaks the 
$SU$(1,1) symmetry, the echoes will not fully recover the initial state. 
This is similar to spin echoes. 
When interactions between spins or other effects break the $SU(2)$ symmetry, 
one uses the spin echo to trace these effects by measuring the imperfect revival.
In this section, we consider that 
a static quartic $r^4$ potential exists as a non-harmonic perturbation in the trapping potential.
To be specific, the Hamiltonians are given by
\begin{align}
   H_1&=(1+\kappa^2)K_0+(1-\kappa^2)K_1+\sum_i a_q r_i^4,\quad nT<t<nT + t_1,\nonumber\\
   H_0&=2K_0+\sum_i a_q r_i^4, \qquad\,\,nT + t_1<t<(n+1)T, \label{Hq}
\end{align}
$a_q$ is made dimensionless with units set by $\omega_0$.
Since there is no simple analytical solution, we numerically solve the GP equation,
\begin{equation}
   i\frac{\partial}{\partial t}\Psi({\bf r},t)=
   \left(-\frac1{2}\nabla^2+\frac1{2}\kappa(t)^2 r^2+a_q r^4+gN|\Psi({\bf r},t)|^2\right)\Psi({\bf r},t),
\end{equation}
where the initial state is prepared as the ground state of $2K_0$ by an imaginary time evolution
such that the system has a period of $2T$ if perfect echoes are delivered. 

\begin{figure}[ht!] 
   \includegraphics [angle=0,width=.75\textwidth]
   {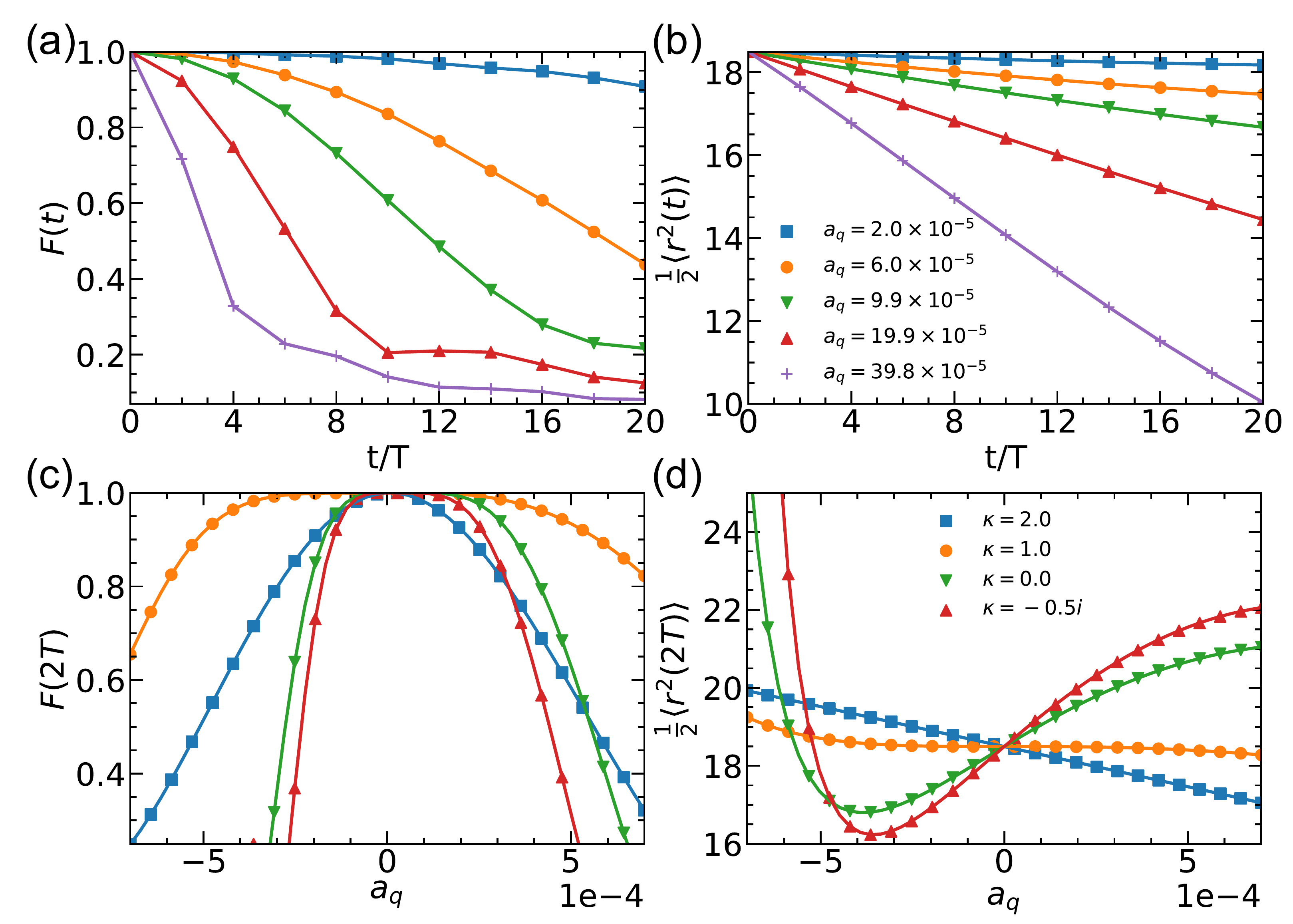}
   \caption
   {  (a-b) Numerical results of $F(t)$ and $\langle r^2(t)\rangle/2$ of 2D BECs
      at stroboscopic time $2nT$ as functions of $n$ for different $a_q$'s.
      $Ng=9600$, $\kappa=2$, $\omega_0=20\times 2\pi {\rm Hz}$ and $t_1=\pi/8$.
      $t_0$ is determined by Eq.(8) in the main text.
      (c-d) Numerical results of $F(t)$ and $\langle r^2(t)\rangle/2$ of 2D BECs
      at $t=2T$ as functions of $a_q$ for different $\kappa$'s.
      $Ng=9600$, $\omega_0=20\times 2\pi {\rm Hz}$ 
      and $t_1=\pi/8$.
      $t_0$ for different $\kappa$ is determined by Eq.(8) in the main text.
   }\label{figs3}
\end{figure}

As shown in Fig.\ref{figs3}(a-b), for a given $a_q$, 
the overlap $F(2nT)$ between the wavefunction at $t=2nT$ and initial state decreases with increasing $n$.
The expectation value of $r^2$ at $2nT$ also deviates from a constant. 
In Fig.\ref{figs3}(c-d), we explicitly show how $F(2T)$ and $\langle r^2(2T)\rangle$ change as functions of $a_q$.  
Thus, these revival signals allow experimentalists to trace the amplitude of the quartic potential. 

\end{document}